\documentclass[9pt,twocolumn,twoside]{osajnl}

\journal{ol} 

\setboolean{shortarticle}{true}

\usepackage{lineno}
\usepackage{afterpage}
\usepackage{everyshi}
\usepackage{caption}
\usepackage[
   justification=justified,
   format=plain]{caption}

\title{Background-penalty-free waveguide enhancement of CARS signal in air-filled anti-resonance hollow-core fiber}

\author[1,*]{Aysan Bahari}
\author[1]{Kyle Sower}
\author[2]{Kai Wang}
\author[1]{Zehua Han}
\author[1]{James Florence}
\author[3]{Yingying Wang}
\author[3]{Shoufei Gao}
\author[4,5]{Ho Wai Howard Lee}
\author[1,6,7]{Marlan Scully}
\author[1,8,9]{Aleksei Zheltikov}
\author[1,6]{Alexei Sokolov}

\affil[1]{Institute for Quantum Science and Engineering, Department of Physics and Astronomy, Texas A\&M University, College Station, Texas 77843, USA}
\affil[2]{School of Physics and Astronomy, Sun Yat-sen University, Zhuhai campus, China}
\affil[3]{Institute of Photonics Technology, Jinan University, Guangzhou 510632, China}
\affil[4]{Department of Physics and Astronomy, University of California, Irvine, CA 92697, USA}
\affil[5]{Beckman Laser Institute and Medical Clinic, University of California, Irvine, California 92697, USA}
\affil[6]{Department of Physics, Baylor University, Waco, Texas 76798, United States}
\affil[7]{Department of Mechanical \& Aerospace Engineering, Princeton University, Princeton, New Jersey 08544, USA}
\affil[8]{Russian Quantum Center, ul. Novaya 100, Skolkovo, Moscow Region, 143025, Russia}
\affil[9]{Physics Department, International Laser Center, M. V. Lomonosov Moscow state University, Moscow, 119992, Russia}
\affil[*]{Corresponding author: a.bahari@tamu.edu}

\begin{abstract}
We study coherent anti-Stokes Raman spectroscopy in air-filled anti-resonance hollow-core photonic crystal fiber, otherwise known as ‘revolver’ fiber. We compare the vibrational coherent anti-Stokes Raman signal of N$_2$, at $\sim$ 2331 cm$^{-1}$, generated in ambient air (no fiber present), with the one generated in a 2.96 cm of a revolver fiber. We show a $\sim$ 170 times enhancement for the signal produced in the fiber, due to an increased interaction path. Remarkably, the N$_2$ signal obtained in the revolver fiber shows near-zero non-resonant background, due to near-zero overlap between the laser field and the fiber cladding. Through our study, we find that the revolver fiber properties make it an ideal candidate for the coherent Raman spectroscopy signal enhancement.
\end{abstract}

\setboolean{displaycopyright}{true}
\DeclareUnicodeCharacter{2212}{-}

\begin{document}

\maketitle


Photonic crystal fibers (PCFs), also known as microstructured fibers, have a great variety of hole arrangements leading to different properties suitable for versatile applications \cite{BROENG1999305}. High harmonic generation, gas-based non-linear optics, and atom and particle guiding are only a few examples of PCF's diverse applicability \cite{Russell358}. Hollow-core photonic crystal fibers (HC-PCFs), a new class of these PCFs where light can be guided in the central hollow channel, are ideal candidates for many linear and non-linear studies due to their flexibility in design, high non-linearity, and tunable dispersion properties \cite{Benabid1999}. 
Aside from the design, pressure inside their hollow core is a parameter that can be tuned to manipulate the dispersion and achieve better phase matching critical to four-wave mixing (FWM) processes such as coherent anti-Stokes Raman scattering (CARS) \cite{Russell2014}. Moreover, due to the HC-PCF's optical confinement and low transmission loss, the distance with constant laser intensity is greatly increased compared to focused laser beams, without the need for optical cavities \cite{Peeters2000,Li:08}.
The CARS spectroscopy in multi-mode fibers  and double-clad fibers has been studied in the past \cite{He1989CoherentRS,Brustlein:11}. For example, a specific double-clad HC-PCF has been used in the CARS experiments to reduce the inherent coherent Raman noise that is generated in the optical fibers \cite{Brustlein:11}. Anti-resonance hollow-core fiber (AR-HCF) with a leaky mode nature is a type of fiber that offers even more advantages over conventional PCFs and HC-PCFs. AR-HCFs with the negative curvature of the core-cladding interface which reduces the optical losses were first proposed in Fiber Optics Research Center of the Russian Academy of Sciences (FORC RAS) in 2011 \cite{Pryamikov11}.  It is important to realize that the cladding of AR-HCFs is not structured as a crystal, so, they are not classified as PCF. The optical properties of this class of fibers are defined only by the reflection of light on the elements of the core-cladding interface. AR-HCF has gained tremendous attention in the fiber optics community due to its low loss properties as well as the moderate cladding fabrication efforts necessary, as compared to many different HC-PCFs \cite{Russell358,Frosz:13,Knight2003}. The broadband single mode operation, high power pulse delivery, and high damage threshold are some of the many advantages of these types of fibers.  By decreasing the coupling between particular waveguide modes in AR-HCF, the anti-resonance phenomenon can significantly reduce optical loss, and hence, improve the quality of the output beam  \cite{Zheltikov2008}. Furthermore, the pressure tuning ability of AR-HCF, as well as its peculiar dispersion behavior around the zero-dispersion wavelength, allows for simultaneous phase-matching of all known Raman transitions of multi-species analytes \cite{Tyumenev:19}. \\
\indent{AR-HCFs provide low-loss broadband guidance even for core diameters as large as tens of micrometers, reducing the pulse energies required for strong nonlinear interactions from the millijoule to the microjoule level, therefore,  allowing scaling from kHz to MHz repetition rates \cite{Kottig17}. Due to this nature of wave-guiding, these classes of fibers are ideal candidates for enhancement of many non-linear processes, by extending the interaction length, and therefore, improving the efficiency of the non-linear processes such as self-phase modulation, super continuum (SC) generation, stimulated Raman scattering, and other FWM process. FWM is an extremely phase-sensitive non-linear process, with its efficiency determined by the relative phases of the four interacting beams and how these phases change over longer distances \cite{PhysRevA2004}.}   
In this work, we use picosecond, narrowband laser pulses at 1064 nm wavelength as our pump and probe, and a fiber-based super-continuum source at 1150 - 1700 nm as Stokes to produce nitrogen CARS signal at $\sim$ 2331 cm$^{-1}$ in a 2.96 cm piece of AR-HCF. Our setup allows simultaneous, multiplex CARS signal detection within the range of Raman shifts from 600 to 3500 cm$^{-1}$. We also study different properties of AR-HCF such as dispersion factor (D) \cite{Lee2020} to predict the wave-vector mismatch ($\Delta$k) in air-filled AR-HCF. The fiber that we use has a broadband guidance that covers the range of the four beams involved in our CARS experiment.  We then show up to 170 $\pm$ 1 times enhancement in nitrogen signal when AR-HCF is used in free-coupling regime compared to the case when no fiber is present. This enhanced signal shows near-zero non-resonant background (NRB), the main limiting factor in CARS spectroscopy \cite{Eesley1981,DRUET19811}. In other words, the fiber’s cladding does not add a substantial NRB to the enhanced signal. This fact is attributed to a near-zero overlap between light, confined in the hollow core, and the fiber's cladding.  Residual NRB produced in the cladding glass (due to a finite overlap) and in air is minimal.

Figure \ref{fig:sem} (a,b) show the schematic and the scanning electron micrograph (SEM) image of our AR-HCF manufactured using a stack-and-draw technique \cite{Ding2020, Gao2016}. We recorded pump and Stokes' beam profiles before and after the AR-HCF (figure \ref{fig:sem}, c-f). This measurement was done when the coupling efficiency for the both pump and Stokes beams  were close to 46$\%$ in a free-coupling regime. Figure \ref{fig:sem} (d,f) show the near-field mode profile of the core-guided modes of the fiber for pump and Stokes beams when the CARS signal was observed. Note that the pump beam is not a perfect Gaussian due to the characteristics of the picosecond Nd:YVO$_4$ laser (APLX-10, Attodyne, Inc.) that we are utilizing in our setup. While the capillary approximation is a convenient way to describe the group velocity dispersion (GVD) and D factor in most HC-PCFs, it does not work for this particular fiber due to its unique structure. However, there are other models that can describe these values for the AR-HCF more accurately. We calculate and plot the dispersion factor of our revolver fiber using  the anti-resonant ring model \cite{zeisberger2018}, which utilizes Zeisberger-Schmidt approximation. In this model, the chromatic dispersion is dominated by the resonances of the annuluses while the order and number of these annuluses has the minor impact on the fundamental core dispersion. The effective refractive index (n$_{eff}$) of a waveguide mode confined within the fiber core can be calculated as:

\vspace{-1.5em}

\begin{figure}[ht]

\centering
\includegraphics[width=3.2in]{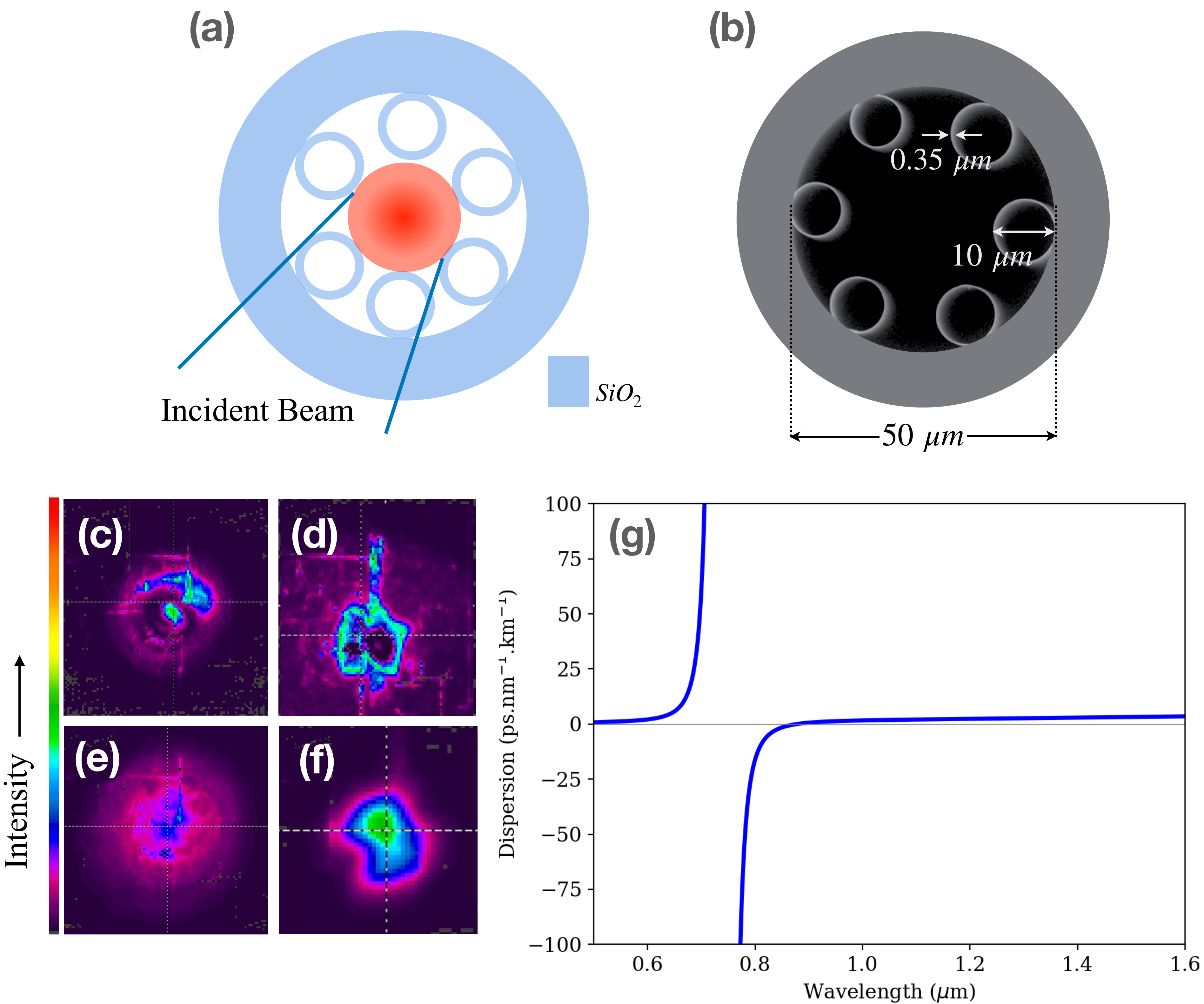}

\caption{Properties of the AR-HCF used in our experiment, (a) schematic of the structure, (b) SEM image of the cross-section and the important measurements (the radius of each ring, the thickness of their walls, and the core size of the fiber), (c,e) before and (d,f) after fiber beam-profiles of pump and Stokes respectively, and (g) the calculated dispersion factor using the ring model. }
\vspace{-1.5em}
\label{fig:sem}

\end{figure}

\begin{equation}
n_{eff}(\lambda)= n_{g}(\lambda)-A {\lambda}^2/n_g(\lambda)-BE(\lambda)C(\lambda){\lambda}^3~,
\end{equation}

\vspace{0.1cm}

\noindent{where $n_g(\lambda)$ is the refractive index of the gas filled in the core of the fiber and A, B, C and E are defined parameters: $A=\frac{j_{01}^2}{8 {\pi}^2 R^2} ,\hspace{0.1cm
} B= \frac{j_{01}^2}{16 {\pi}^3 R^3}$, $C(\lambda)=\cot{\frac{2\pi w \sqrt{n_s^2(\lambda)-n_g^2(\lambda)}}{\lambda} }$, and $E(\lambda)=\frac{n_s^2(\lambda)-n_g^2(\lambda)}{n_g^3(\lambda) \sqrt{n_s^2(\lambda)-n_g^2(\lambda)}}$ for the fundamental mode (HE$_{11})$ with R being the core radius, $n_s$ being the refractive index of the rings' material (glass), and $j_{01}=2.405$ being the root of the zeroth order Bessel function (j$_0(u)=0$) 
\cite{zeisberger2018}. Here, $n_{eff}$ is used to calculate the propagation constant, and therefore, D factor. Due to the cotangent dependence in $C(\lambda)$, the zero dispersion wavelengths (ZDWs) can be tuned by the capillary thickness $w$ (in our case 0.35 $\mu$m) to optimize the desirable processes at the operating wavelengths. However, if the process of interest spans a broad wavelength range, it is likely to cross a resonance, making simple approximations of the phase mismatch difficult. A Taylor expansion of the propagation constants will fail due to singularities. The resonant wavelengths are given by the wavelengths where the cotangent diverges:}
\vspace{-0.1cm}

\begin{equation}
\lambda_{ZD} = \dfrac{2w}{l}\sqrt{n_s^2-n_g^2}~,
\end{equation}

\vspace{-0.1cm}
\noindent{where $l$ = 1,2,... . The cotangent dependence also produces broad regions of anomalous dispersion, which can be balanced against the normal dispersion of the medium inside the fiber to provide another avenue for phase matching. Figure \ref{fig:sem} (g) shows the dispersion factor of our AR-HCF calculated using these considerations which indicates no singularities in the range of our interest ($\sim$ 800 nm-1700 nm). Applying the phase matching condition in CARS process: $\Delta k= k_{Stokes} + k_{signal} -2 k_{pump}$ where $k_{pump} = k_{probe}$ (degenerate CARS), we can have a decent estimation of the phase mismatching as a function of pressure and wavelength \cite{Bahari:19} for our experiment giving the possibility of using pressure-controlled setup which uses different buffer gases to enhance CARS spectroscopy even further. }

\begin{figure}[ht]
\centering
\vspace{-1em}

\includegraphics[width=3.4in]{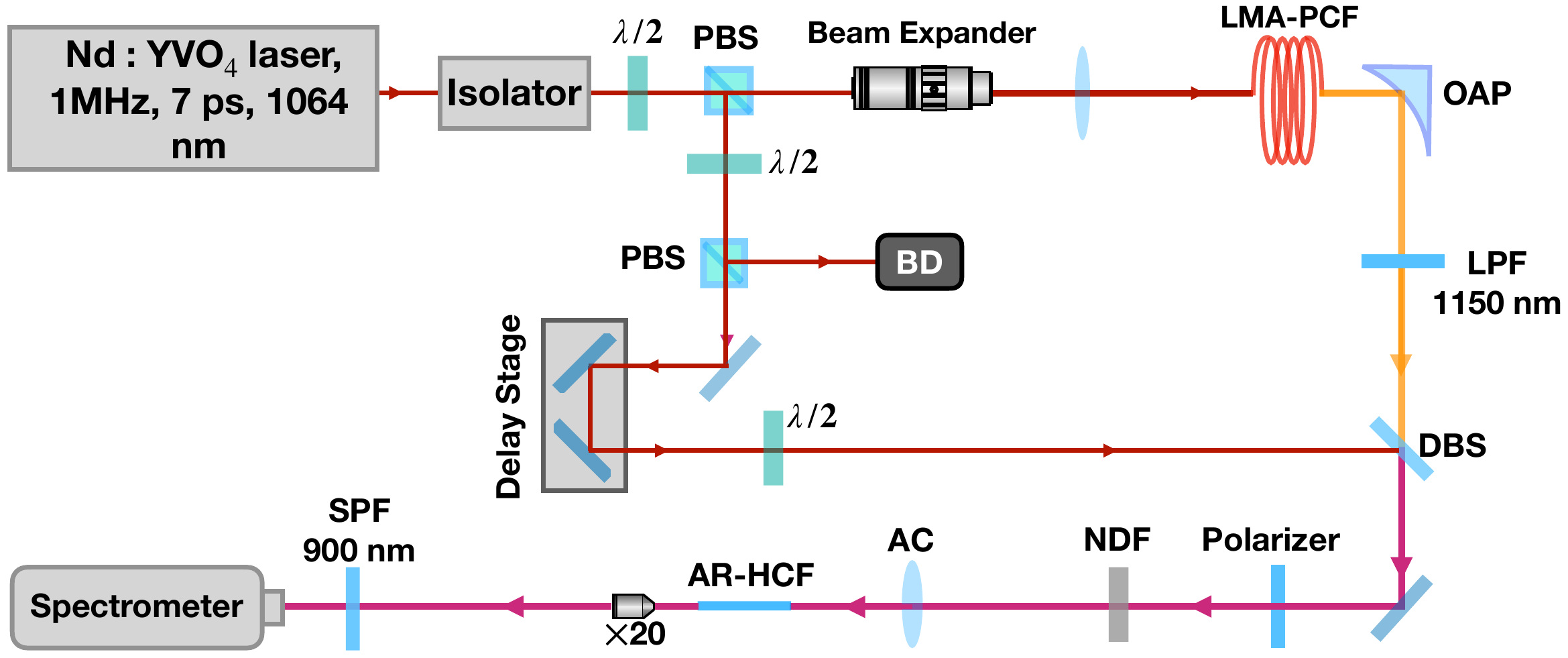}
\caption{Schematic of our experimental wave-guided CARS setup. A long-pass filter at 1150 nm is used to cut the lower side of the SC spectrum and a short-pass filter is used right before the spectrometer to get rid of the intense bands of the fundamental and the SC. Abbreviation for the optical elements: PBS, polarizing beam splitter; BD, beam dump; $\lambda /2$, half wave-plate; OAP, off-axis parabola; LMA-PCF, large mode area photonic crystal fiber; LPF, long-pass filter; DBS, dichroic beam splitter; NDF, neutral density filter; AC, achromatic lens; AR-HCF, anti-resonant hollow-core fiber; and SPF, short-pass filter. 
}
\vspace{-1em}
\label{setup}
\end{figure}

Our setup is shown in Figure \ref{setup}. We use a laser beam generated in a Nd:YVO$_4$ laser with 1 MHz repetition rate, 7 ps pulse duration, and the wavelength of 1064 nm divided into two legs using a polarizing beam splitter (PBS). Half-wave plates are used in combination with each PBS to control the power of each leg. One leg is used as our pump and probe, and the other leg goes through an approximately 2 m long large mode area  fiber (LMA-20) to generate a broadband SC (900 – 1700 nm) using a plano-convex lens (LA1509-C, Thorlabs, Inc.). The generated SC is then collimated by an off-axis parabola (OAP) (MPD129-P01, Thorlabs, Inc.) and is guided to the rest of the setup.  A long-pass filter (FELH1150, Thorlabs, Inc.) is used right after the OAP to cut the lower side of the SC spectrum at 1150 nm to get rid of the strong fundamental at 1064 nm while detecting the relatively weak CARS signal on the spectrometer.

The both beams then are combined on a dichroic beam splitter (LPD02-1064RU-25, Semrock, Inc.) and aligned to a polarizer and a neutral density filter (NDF) to adjust their powers, and then are focused into a 2.96 cm long AR-HCF with an infinity corrected achromatic lens (AC254-050-C, Thorlabs, Inc.) with a focal length of 5cm. The fiber is cut to this length using a high precision, large diameter optical fiber cutter/cleaver (3SAE-LFC II, Technologies, Inc.) and then, is mounted on a fiber holder (HFF001, Thorlabs, Inc.) fixed on a manual 3-axis translation stage for a precise alignment. A microscope objective (M-20 x, Newport, Inc.) is used after the fiber to collect and collimate the output beam into a short-pass filter with cutt-off wavelength at 900 nm (FES0900, Thorlabs, Inc.) and then a CCD spectrometer (Holospec, Andor with the attached TE-cooled CCD (iDus416, Andor, Inc)). Argon, neon, and mercury lamps (StellarNet Inc.) are used to calibrate each CCD channel separately. The delay stage in the setup controls the temporal overlap between pump and Stokes beams. To calibrate our setup further, we optimize a CARS signal of a sample of polystyrene beads, with bead size of  3 $\mu$m (No. 77523, SigmaAldrich, Inc.) deposited on a coverslip (Micro Cover Glasses, VWR) under the microspectrometer \cite{Shen2018}.

\begin{figure}[ht]
\centering
\includegraphics[width=3.2 in]{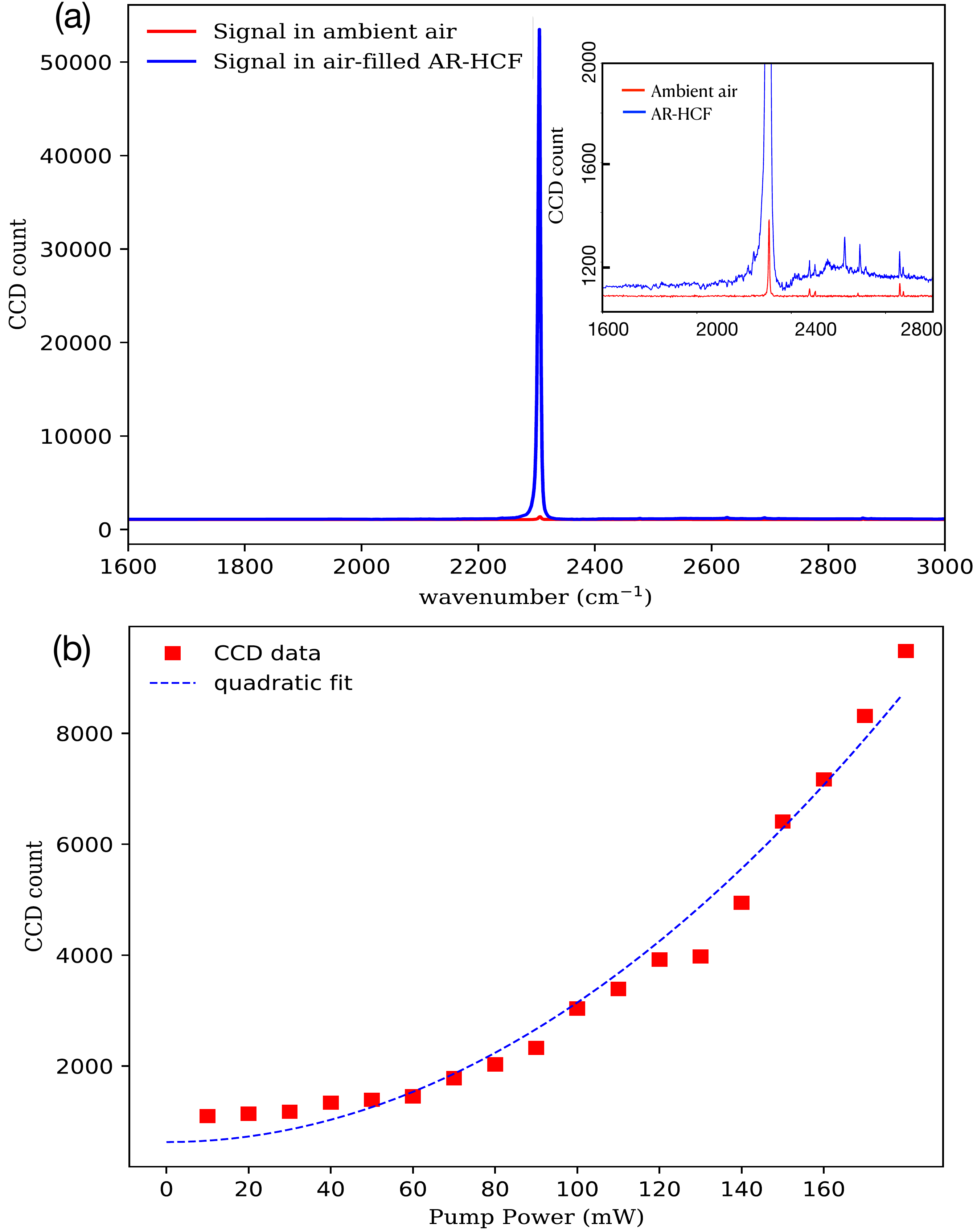}
\caption{ (a) The CARS signal of nitrogen at 2331 cm$^{−1}$ (Q-branch vibrational transition) in a 2.96 cm long air- filled AR-HCF compared to the CARS signal from the ambient air (no fiber present). Inset: enlarged plot in the lower CCD count region for better resolution. The spikes in the region between 2400 cm$^{-1}$ and 2800 cm$^{-1}$ only appear at very short acquisition times (here 0.05s).  This is because of the way that the CCD spectrometer operates. (b) Pump power dependent CARS signal measurement. Since the pump beam is used as both pump and the probe beams, the quadratic dependence of the CARS signal is expected.}

\label{results}
\end{figure}

Our final results are shown in figure \ref{results}. The top panel (a) compares the two nitrogen CARS signals obtained one from the free-coupling pump and Stokes in a 2.96 cm long air-filled AR-HCF (the blue curve) using an AC lens with 5cm focal length and the other (red curve) from ambient air, meaning no fiber was present and pump and Stokes were simply overlapped and focused in the air using the same AC lens. The CARS signal collected from AR-HCF is measured on CCD while the coupling efficiencies for the pump beam is around 70$\%$ and for the Stokes beam is 65$\pm$5$\%$. These values are  calculated by simply dividing the measured output powers by the measured input powers. The coupling efficiency for multimode fibers can be evaluated more precisely using the overlapping integral theory \cite{NIU2007315}. The input powers of pump and Stokes beams are 180 mW and 3.5 mW, respectively.  

\vspace{-0.3cm}
Comparing these two curves shows 170 $\pm$ 1 times enhancement in the collected signal from the AR-HCF. This enhancement was possible by providing a longer interaction path through the HCF, and therefore, improving the efficiency of the phase-matched FWM. We can also calculate the enhancement factor theoretically considering
$P_{CARS} = I_{CARS} \times S$, where P, I, and S stand for power, intensity, and the interaction cross-section area respectively. Considering that the beam size before the focus is 1 cm, focal length of the AC is 5 cm, the AR's core radius is 15 $\mu m$, and $I_{CARS} \propto P^2_{Pump} P_{Stokes}  l^2/S^2 $ ($l$ being the interaction length), we can calculate the enhancement factor $\eta$ as:

\begin{equation}
\eta = \frac{(l_{AR}/S_{AR})^2 T^2_{pump} T_{Stokes}}{(l_{ambient}/S_{ambient})^2}   .
\end{equation}

Here, $l_{AR}$ is the length of the fiber, $S_{AR} = \pi a^2$ ($a$ being AR's core radius), $l_{ambient}$ is two times the Rayleigh range in ambient air, and $S_{ambient}$ is equal to $\pi w^2_0$ with $w_0$ being the beam waist at the focus in ambient air. T$_{pump}$ and T$_{Stokes}$ are the coupling efficiencies for pump and Stokes beams, respectively. We find the theoretical enhancement factor to be around 170 which is perfectly matching with what we obtain experimentally.

Moreover, no substantial NRB is added to the signal from the fiber's cladding due to its unique design properties. This fact is due to a near-zero overlap between light, confined in the core, and the fiber’s cladding. As one can see, the residual non-resonant background, produced in the glass cladding (due to a finite overlap) and in air is minimal. Our measured CARS  signal to NRB ratio in the air-filled AR-HCF is $\sim$ 350 while the same ratio that we measure for ambient air is $\sim$ 520 at its best. 
There are several techniques for NRB reduction in CARS spectroscopy such as time-resolved scheme \cite{Selm:10,Volkmer2002}, heterodyne detection \cite{Kee,Potma2006}, and polarization control \cite{Song1976,Bunkin1977,Akhmanov1978,Koroteev,Oudar1979}. The latter can be achieved through polarization control of the involved beams.  In principle, if further background suppression is needed in our future experiments, we can utilize the polarization technique, since anti-resonance hollow-core fibers can be sufficiently short and straight and be designed to maintain polarization of guided fields, thus enabling polarization NRB suppression for waveguide-enhanced CARS.\\ 
\indent{Note that in both cases the focal length and the beam size at the focus are the same. The reason that the oxygen CARS signal  at 1556 cm $^{-1}$ was not detected is because of the efficiency roll-off of the CCD spectrometer at the edge of each channel (channel 1: 600 - 1500 cm$^{-1}$ and channel 2: 1500 - 3500 cm$^{-1}$). Moreover, regardless of the fact that the oxygen signal should be 15 times weaker than the nitrogen one \cite{Lucht:87}, the use of a SPF at 900 nm to get rid of the strong fundamental and SC filters the oxygen signal even more.}

Finally, although the primary enhancement comes from the elongated interaction length provided with the hollow-core of the fiber, for the ultra-short pulses in such non-linear processes, temporal walk-off due to dispersion in the medium is unavoidable \cite{Boyd}. To overcome this obstacle, pressure tuning inside the core is an option to manipulate the dispersion and fulfill the phase matching condition \cite{PhysRevA2004,Trabold2017}.
In figure \ref{results} (b), as a proof of concept, we demonstrate the nitrogen CARS signal (peak count) dependence to the power of pump beam which is also used as probe, therefore, a quadratic dependence is expected and measured experimentally in this plot.

We successfully demonstrate $\sim$ 170 times enhancement with a near-zero NRB  in nitrogen CARS signal using a short piece of AR-HCF in a free-coupling regime. This method, which utilizes a much less complicated table-top setup compared to most fiber-based experiments, can be extended to many multi-spices gas samples and also to biomedical solutions and functionalized residues. In the near future, we are planning to expand this method to detecting the CARS signal of liquids such as different types of protein solutions and antibodies.

\begin{backmatter}
\bmsection{Funding}
This work is supported by the Robert A. Welch Foundation (Grant No. A-1261 and A-1547) and King Abdulaziz City for Science and Technology (KACST).

\bmsection{Acknowledgments}
A. B. and Z. H. are supported by the Herman F. Heep and Minnie Belle Heep Texas AM University Endowed Fund held/administered by the Texas A\&M Foundation. K. S. is supported by the TAMU/Association of Former Students Graduate Merit Fellowship. H. W. H. L. acknowledges support from AFOSR (FA9550-21-1-02204).

\bmsection{Disclosures} The authors declare no conflicts of interest.
\end{backmatter}


\bibliography{sample}


\end{document}